\documentclass[10pt, conference, compsocconf]{IEEEtran}

\ifCLASSINFOpdf

\else

\fi

\makeatletter
\def\ifundefined{\@ifundefined}
\makeatother

\usepackage{setspace}
\usepackage{algcompatible}
\usepackage{algorithmicx}
\usepackage{algpseudocode}
\usepackage{algorithm}
\usepackage{multirow}

\usepackage{url}

\usepackage{srcltx}
\usepackage{amsmath}
\usepackage{amssymb}
\usepackage{graphics}
\usepackage{graphicx}
\usepackage{cite}
\usepackage{psfrag}
\usepackage{epsfig}
\usepackage{subfigure}
\usepackage{amsthm}
\usepackage{thmtools}
\usepackage{color}
\usepackage[justification=centering]{caption}
\usepackage{refstyle}
\usepackage{amsmath}
\usepackage{amsthm}
\usepackage[justification=justified,singlelinecheck=false]{caption}

\usepackage{lipsum}

\usepackage{tabularx}
\usepackage{caption}

\usepackage{array,booktabs,arydshln,xcolor}

\usepackage[T1]{fontenc}      
\usepackage[utf8]{inputenc}   
\usepackage{caption}          
\usepackage{mathtools}        
\usepackage{%
   multirow,                   
   tabu                        
}
\usepackage{xcolor}           
\usepackage{siunitx}
\usepackage{float}
\usepackage{arydshln}
\usepackage{booktabs}
\usepackage{lipsum}
\usepackage{makecell}
\setcellgapes{4pt}

\usepackage{adjustbox}
\usepackage{graphicx}

\usepackage{setspace}

\begin{document}

\title{Risk Assessment of Autonomous Vehicles \\ Using Bayesian Defense Graphs}

\author{\IEEEauthorblockN{Ali Behfarnia and Ali Eslami}
\IEEEauthorblockA{Department of Electrical Engineering and Computer Science \\
Wichita State University, Wichita, KS, USA\\
Email: axbehfarnia@shockers.wichita.edu, ali.eslami@wichita.edu}
}
\maketitle

\begin{abstract}
Recent developments have made autonomous vehicles (AVs) closer to hitting our roads. However, their security is still a major concern among drivers as well as manufacturers. Although some work has been done to identify threats and possible solutions, a theoretical framework is needed to measure the security of AVs. In this paper, a simple security model based on defense graphs is proposed to quantitatively assess the likelihood of threats on components of an AV in the presence of available countermeasures. A Bayesian network (BN) analysis is then applied to obtain the associated security risk. In a case study, the model and the analysis are studied for GPS spoofing attacks to demonstrate the effectiveness of the proposed approach for a highly vulnerable component. 



\end{abstract}


\IEEEpeerreviewmaketitle

\section{Introduction}
An autonomous vehicle (AV) is able to perceive its environment, navigate, and maneuver without human action. 
AVs, unlike traditional vehicles, rely solely on sensors, processing systems, and communication messages for making driving decisions. This very large amount of sensing and data processing creates opportunities for adversaries to compromise vulnerable components in AVs. AVs will be regularly used only if their security level is higher than a predefined threshold. Therefore, it is vital to recognize threats, classify them, and develop protection strategies for AVs. Protection solutions must eventually result in quantitative measurements to assure AV reliability.

In recent years, experts have continuously sought to identify gaps towards improving the security of AVs. Some researchers \cite{Petit_Survey, Survey17, yan16} have studied potential cyberattacks and their implications on automated and cooperative AVs. In particular, Petit and Shladover \cite{Petit_Survey} categorized threats as high, medium, and low, based on some criteria used in \cite{FMEA03}, such as the feasibility of attack, the probability of attack success, etc. In order to evaluate countermeasures on AVs, Petit et. al \cite{petit2015remote} applied some redundancies and optic materials.
While this work and similar studies are crucial to identify research gaps and possible solutions, they have not provided a unified platform for security measurement in the presence of anti-attack techniques.
On the other hand, researchers have widely employed attack and defense graphs as powerful tools to analyze computer networks' security.
An attack graph is a graphical representation of all paths through a system that end in a state where an intruder successfully exploits the system.
A defense graph, as explained later, is a mitigation mechanism which is formed similar to an attack graph, with the only difference that the leaf nodes are countermeasures \cite{kordy14}. Several authors in \cite{roy12, kordy10} introduced countermeasure and attack-defense trees as graphical models to study the security of systems using probabilistic analysis. In spite of such efforts, there is no platform based on defense graphs to measure the likelihood of threats and risks for vulnerable components in AVs.


In this paper, we take a novel yet simple approach using the defense graph concept to address the existing gap for quantitative security assessment in AVs.
Our main contributions can be summarized as follows:
%
\begin{itemize}
\item We propose a plain security model in which vulnerable components can be monitored through their security states. These states together represent the security state of an AV.

\item We employ a defense graph as a security model, and then evaluate it based on prominent risk assessment models such as EVITA (E-safety vehicle intrusion protected applications) in order to study the effect of countermeasures.

\item We derive a Bayesian defense graph for detecting fake GPS signals in the presence of anti-spoofing techniques. Using probabilistic inference, we demonstrate that threat likelihoods of less than $0.01 \%$ can be reached using a set of protection techniques.

\end{itemize}

The rest of this paper is organized as follows. Section II explains the proposed model, threat identification and risk assessment, and BN model in the presence of uncertainties in forming a defense graph.
Section III applies the proposed model to the GPS unit as a highly vulnerable component in AVs. Various combinations of GPS anti-spoofing techniques are considered towards measuring the protection levels provided by them collectively. Section IV concludes the paper.

\section{Modeling of Secure Autonomous Vehicles Using Bayesian Networks}
In this section, we provide a theoretical model to measure the security of AVs.
First, we describe a security model based on defense graphs for monitoring vulnerable components in an AV. Then, we explain how we consider threats and risk assessment for the model. Finally, we apply BN analysis as a simple but powerful tool to perform security measurements in this model.


\subsection{Proposed Security Model}
\begin{figure}
    \centering
    \subfigure[]
    {
        \includegraphics[width =2.2 in , height=1.2 in]{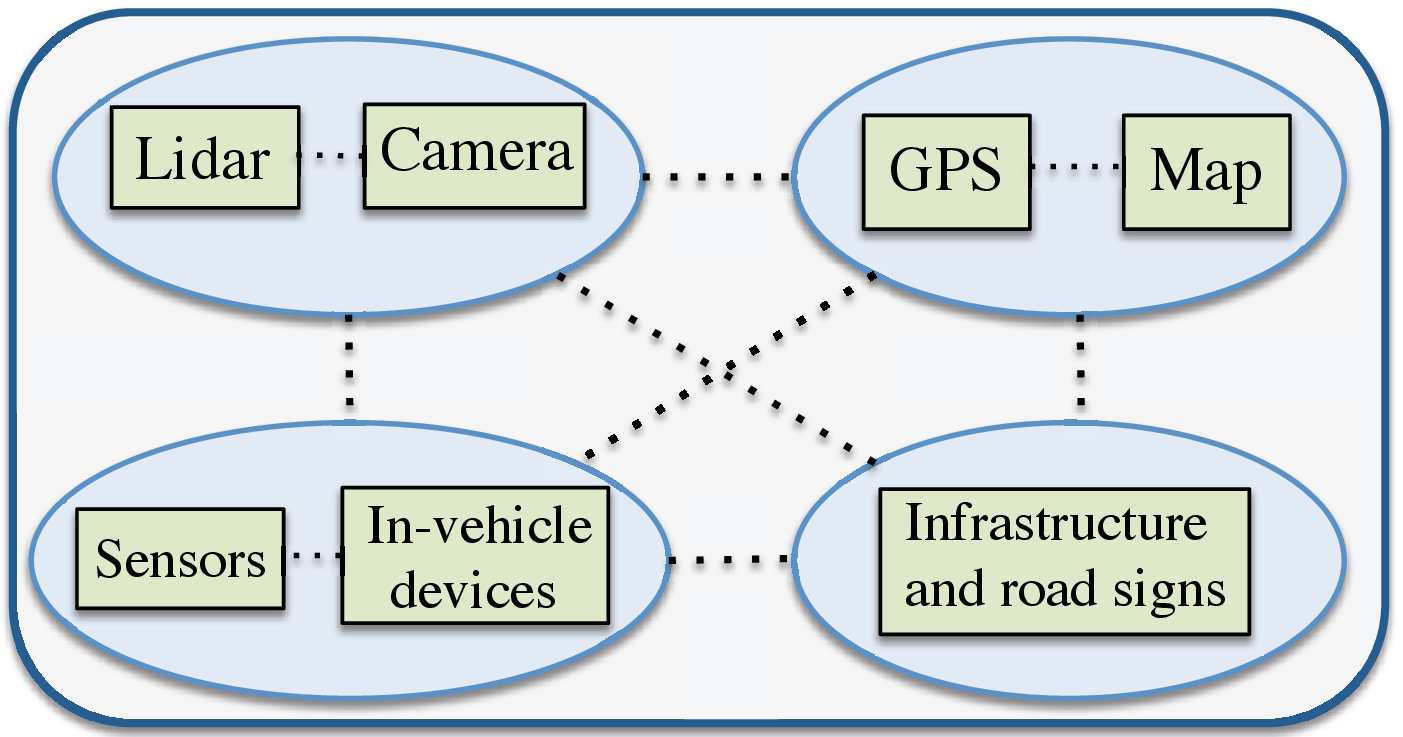}
        \label{fig:E_b}
    }
    \\
    \subfigure[]
    {
        \includegraphics[width = 3.4 in , height = 3.0 in]{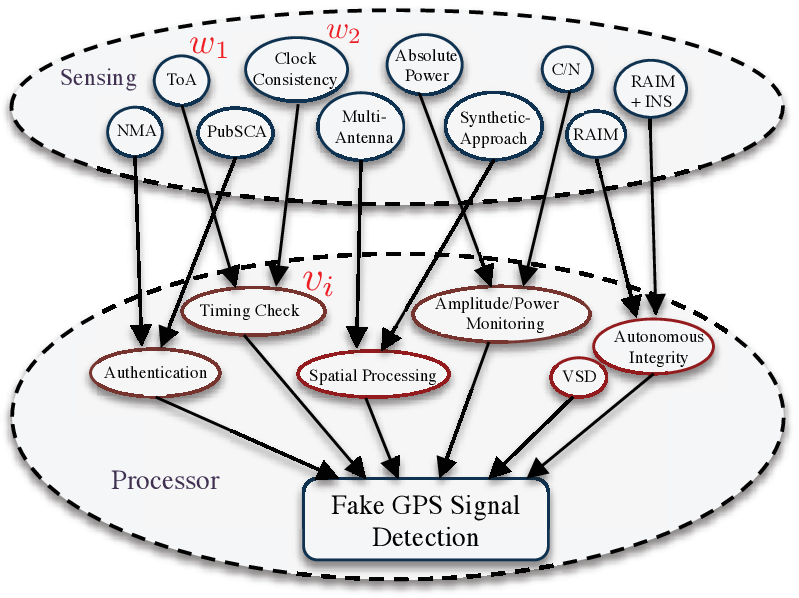}
        \label{fig:E_c}
    }
    \caption{ (a) Security monitoring unit for AV, and (b) graphical model for a secure GPS component in AV.}
    \label{fig:EF}
  \vspace{-0.5cm}
\end{figure}

A security monitoring unit is an essential part of an AVs' central processor, which investigates all required data to assure the security of a vehicle. Fig. \ref{fig:E_b} shows a typical security monitoring unit consisting of major attack surfaces \cite{Petit_Survey}. Here, each attack surface is referred to as ``component''. As a part of processor, this unit has access to all required data for protection purposes.

In order to monitor the security status of an AV, we assess all vulnerable components. 
Let us define SV as the security state of an AV as follows:
\begin{align}\label{States}
SV \triangleq \{S_1, S_2, ... , S_n\},
\end{align}
where $S_i$ denotes the security state of the $i$th vulnerable component. Each security state could be either normal or abnormal. A component is in an abnormal state when an attacker successfully mounts an attack on the component (i.e., the component is exploited). To ensure security, we could employ countermeasures for vulnerable components to prevent them from being exploited. Consideirng this point, we define a set of defense techniques as observable contexts to determine the security states as follows:
\begin{align}\label{Context}
S_i = f(C_{i1}, C_{i2}, ... , C_{ik}).
\end{align}
%
Each observable context $C_{ij}$ refers to the $j$th element of an defense technique related to the $i$th vulnerable component. To clarify this, consider Fig. \ref{fig:E_c} as a graphical representation model for protecting a GPS component. Hence, this graph can be considered as a defense graph. As shown, we employ several techniques such as a timing check ($v_i$) in the processor to detect counterfeit GPS signals. Each technique needs some elements, such as clock consistency ($w_1$), to be accomplished. These predefined elements as part of defense techniques provide observable contexts ($C_{ij}$'s). 
We utilize information from observable contexts and apply Bayesian inference as a mathematical reasoning method to characterize unobservable security states ($S_i$'s). In the following section, we discuss threats against $C_{ij}$'s and the risk assessment of $S_i$'s. 


%

\subsection{Threat Identification and Risk Assessment}\label{sec:risk}
Threat identification is the first step towards devising a security model for a system. In this paper, we assume that vulnerable components of an AV have been already identified, thanks to previous works such as \cite{Petit_Survey, Survey17, yan16}. This allows us to employ defense graphs formed by countermeasures to protect AVs. Threats in the context of a defense graph could be interpreted as possible ways that counterfeit signals could go through the countermeasures without being detected. This means that a vulnerable component can be successfully exploited if none of corresponding countermeasures detect the fake signal. 
For instance, if an attacker remains undetected by the authentication countermeasure in Fig. \ref{fig:EF}(b), it might be able to tamper with GPS information, causing a major threat.
There exist several frameworks such as Microsoft's STRIDE (Spoofing, Tampering, Repudiation, Information disclosure, Denial of service, and Elevation of privilege) for threat identification that have been demonstrated to work well for AVs \cite{sahara}. 


Once threats are identified, risk assessment could be carried out to determine the level of security in a system. 
There exist some methodologies to do the risk assessment, such as EVITA and CVSS (Common Vulnerability Scoring System). Risk assessment contains two fundamental parts: impact (or severity) of threats, and likelihood of threats. In order to estimate the impact of a threat, one could employ parameters that directly associate with harm to stakeholders. Safety, privacy of drivers, operational performance, and financial losses of a vehicle are four factors commonly used in automative risk models \cite{islam16, Hu09}. The level of each factor can be categorized as none, low, medium, and high. To approximate the likelihood of a threat, one should calculate the probability of a successful attack. This could also be evaluated based on the above risk assessment models. For instance, expertise, knowledge of target, window of opportunity (including time requirement), and equipment are four main parameters in EVITA to estimate the likelihood of threats. The level of each can be rated between $0$ to $3$. Table \ref{foo} shows examples of evaluation of impact and likelihood of threats.

Having the levels of impact and likelihood, we can compute the risk which is a function of both. A standard risk model can be defined as follows:
\begin{align} 
Risk = Likelihood \times Impact \label{Risk_std}
\end{align}
where risk indicates risk for a set of countermeasures. 
The effect of countermeasures appears only in the likelihood, and not the impact, of a threat. Therefore, countermeasures directly affect the value of likelihood, while impact is determined by the functionality of a component (such GPS) within the AV.
Also, the intrinsic uncertainty of attacks leads us to assess parameters based on probabilities. 
Here, $impact$ can be directly estimated from the parameters in risk rating methodologies, such as Table I(a). To obtain $likelihood$ for a component, however, two points should be considered: i) the quantity and the quality of the employed countermeasures and ii) cause-effect relationships between the elements of countermeasures. The former can be captured through standard parameters (e.g. Table I.(b)), and the later can be represented by directed acyclic graph (DAG) as a defense graph. Having the graph with related parameters enable us to infer likelihood using BN analysis.

\begin{table}[t]
\centering
\captionsetup{justification=centering}
\caption{ Example of EVITA risk assessment factors: (a) Impact of an attack on GPS, (b) Likelihood of a threat against the ToA countermeasure in Fig. \ref{fig:EF}.} 
{\includegraphics[width =3.5 in , height=1.5 in]{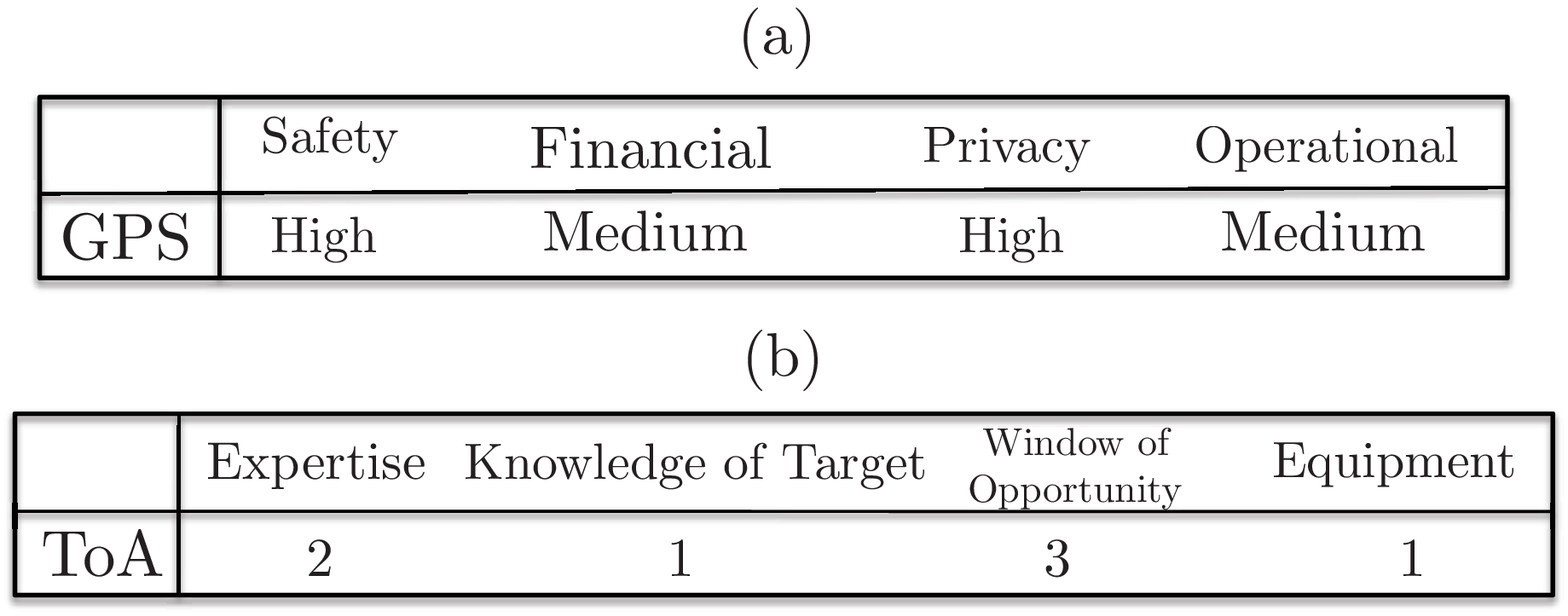}}\label{foo}
\end{table}



\subsection{Bayesian Network and Uncertainty}\label{Uncertainties}

A Bayesian network is a graphical model for probabilistic inference that denotes the relationship between a set of variables by a directed acyclic graph (DAG). A BN is a pair $(S,P)$, where $S$ denotes a network structure, and $P$ denotes a set of conditional probability distributions. Let us consider a DAG $S=(\mathbf{V}, \mathbf{E})$, where $\mathbf{V}=\{v_1, v_2, ...., v_n\}$ represents a set of nodes, and $\mathbf{E}=\{e_1, e_2, ...., e_n\}$ represents a set of edges. Using this definition, each node could denote a countermeasure technique, such as time checking in Fig. \ref{fig:E_c}, or an element of it, such as clock consistency. A directed edge exists from node $v_i$ to node $v_j$, only if there is the possibility for an exploit to be instantiated from $v_i$ to $v_j$. Generally, in order to build a defense graph, the functionality of each node as well as cause-effect relationships between nodes (w.r.t. an application) must be captured in the BN framework. 


Once we build a BN, we are able to perform probabilistic inference. 
Here, we are interested in applying marginal and posterior probability distributions to measure vulnerability for components. 
To clarify this, assume that we want to quantitatively measure the vulnerability of $v_i$ that is shown in Fig. \ref{fig:E_c}. 
Assuming $\mathbf{W}=\{w_1, w_2\}$ as parent nodes of $v_i$, we can compute the following:
%
%
%
%
\begin{align}\label{Marg}
 p(v_i) & = \sum  p(v_i | w_1,w_2 ) \ p(w_1,w_2),             
\end{align}
\begin{align}\label{Post}
p(v_i | \mathbf{W}) \propto p( \mathbf{W} | v_i) p(v_i).
\end{align}
Equation (\ref{Marg}) is a marginal probability distribution to obtain prior probability for $v_i$, and equation (\ref{Post}) represents a posterior probability distribution using the prior probability and $p( \mathbf{W} | v_i)$ as a likelihood distribution. 
Using equations (\ref{Marg}) and (\ref{Post}), we are able to calculate the likelihood of a successful attack on $v_i$, given the vulnerability of $\mathbf{W}$ ($v_i$'s parent nodes). 

\begin{figure}[t]
\centering
\captionsetup{justification=centering}
{\includegraphics[width =3.3 in , height=1.5 in]{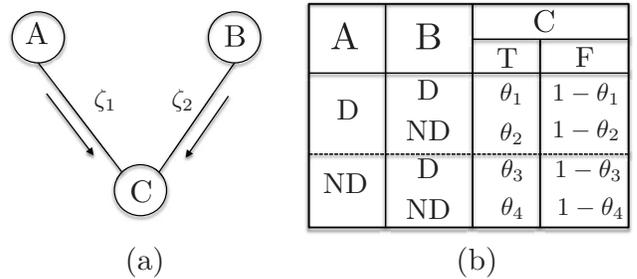}}
\caption{(a) portion of a defense graph, and (b) corresponding conditional probability table.}
\label{fig:B} \vspace{-.6cm}
\end{figure}

Before applying the BN theory to obtain the security state of each vulnerable component, we need to capture uncertainties related to a realistic AV application.
To this end, let us assume that Fig. \ref{fig:B}(a) shows a portion of a complete BN. Each node represents an anti-attack element to protect a vulnerable component. For instance, let us assume nodes $A$ and $B$ are two anti-attack elements for a secure component $C$. To yield a successful attack on node $C$, nodes $A$ and $B$ must have been unable to detect the attack. Hence, the framework for the reasoning of our defense graph is AND logic. Fig. \ref{fig:B}(b) indicates a conditional probability table (CPT) in which different scenarios of detection (D) and not detection (ND) are considered. The true (or false) state signifies a successful (or unsuccessful) detection on a component, respectively. 

\newcolumntype{C}[1]{>{\centering\arraybackslash}p{#1}}
 \renewcommand{\arraystretch}{1}
\begin{table*}[t]
\centering
\captionsetup{justification=centering}
\caption{ Prior probabilities of anti-spoofing techniques for detecting fake GPS signals using EVITA and CVSS.} 
{\includegraphics[width = 7 in , height=.95 in]{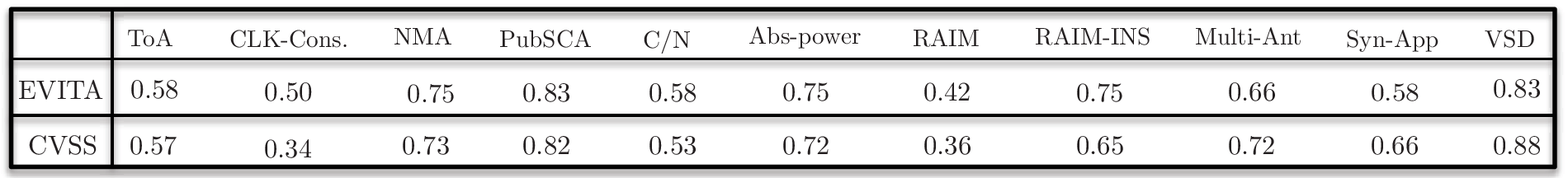}}\label{fii}
\end{table*}

Here, we also account for the uncertainty between neighboring nodes due to their imperfect accuracy and trustworthiness. 
In addition, there exists an inherent uncertainty in attack structures. That is, even though an attack is successfully mounted on nodes $A$ and $B$, there is no guarantee for the attacker to successfully carry out its attack on node $C$. To capture these points, we consider coefficients $\zeta_1$ and $\zeta_2$ between nodes, as shown in Fig. \ref{fig:B}(a). Considering these coefficients, we define $\theta_i$ in the CPT to indicate the probability of a true state in node $C$. In a defense graph, it is reasonable to have a high reliability between nodes, which implies small values for $\theta_i$, $i=1, 2, 3$, and a value close to 1 for $\theta_4$.

In the next section, we investigate the security measurement of GPS signals as a vulnerable component. Other vulnerable components in an AV (e.g., LiDAR, camera) could be investigated in the same fashion.

\section{Case Study: Secure GPS Component} \label{Techniques}
GPS spoofing is among the highest threats for AVs. Hence, in this case study, we investigate the security measurement of GPS using the proposed BN model. In particular, we would like to obtain likelihood and risk for a defense graph shown in Fig. \ref{fig:EF}(b).

\subsection{Modeling and Parameterizing}\label{sec:Case_M&P}
A principle objective of this work is to quantify the security of a GPS component for AVs, by means of the following: (a) building a defense graph using BN model, and (b) parameterizing elements of the graph. Combining these two allows us to make an inference for likelihood, hence risk.

In order to model a defense graph for a GPS component, all possible ways to detect counterfeit GPS signals must be considered. Here, six most effective anti-spoofing techniques are selected. Each technique includes different elements for sensing abnormalities. Fig. \ref{fig:E_c} shows a defense BN model for a GPS component obtained from cause-and-effect relationships among the elements of anti-spoofing techniques. These techniques are well studied in \cite{ReviewGPS,EPFL08, Mul-Ant15, part2,VSD08}.
As can be seen, each technique (e.g., timing check) contains a few elements (e.g., clock consistency) to sense environment and send the required data for processing purposes. However, there is a possibility for an attacker to defeat an anti-spoofing technique which leads us to likelihood.

\begin{table}[t]
\centering
\captionsetup{justification=centering}
\caption{ Example of conditional probability table}
{\includegraphics[width =2.4 in , height=1.4 in]{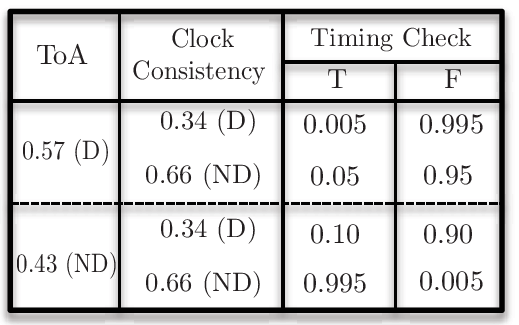}}\label{fee}
\vspace{-0.2 cm}
\end{table} 

To find the value of likelihood, we need to determine the prior probability of each element and the conditional probability between the elements in the graph. We employ three approaches to make these evaluations: (a) EVITA as a risk assessment model, (b) CVSS that uses existing databases such as the National Vulnerability Database (NVD), and (c) several studies that have already addressed similar issues (e.g., \cite{ReviewGPS,EPFL08,warner03, psiaki16}). We apply the first two to find the prior probability and the last one to find the conditional probability. As we mentioned in section \ref{sec:risk}, we use four parameters for EVITA evaluation. For instance, as can be seen in Table I(b), since the summation of values is $7$ and the total possible value is $12$, we derive $\frac{7}{12}$ as the probability of detection for ToA. In CVSS, we consider two major concepts in calculating the scores: the base score (BS) and the temporal score (TS). The BS quantifies the intrinsic attribute of each vulnerability, which is independent of time and user environment. The TS, however, assesses the vulnerability based on properties that might change over time. Using BS and TS scores, the CVSS generates a value from $0$ to $10$ that can be simply converted to a probability by dividing the score over $10$ \cite{frigault08}. Table \ref{fii} indicates the values of prior probabilities based on EVITA and CVSS. To obtain conditional probabilities between graph nodes, we use previous literature to consider all dependencies between anti-spoofing elements. We define four discrete probability levels w.r.t. the accuracy of anti-spoofing methods: $0.995$ (almost sure), $0.99$ (probable), $0.95$ (highly expected), and $0.90$ (expected). These values represent $\theta_i$s in the CPT table of Fig. \ref{fig:B}(b). For instance, Table \ref{fee} shows a CPT using CVSS for the timing check unit. CPTs for the rest of the anti-spoofing techniques can be obtained in the same fashion. Having a BN graphical model and its corresponding CPTs, the next step is to perform an inference to find likelihood for the GPS component.

\begin{table*}[t]
\centering
\captionsetup{justification=centering}
\caption{Likelihood of threats and risk probabilities for a sample of combinations of GPS anti-spoofing techniques. }
{\includegraphics[width = 7 in , height=2.6 in]{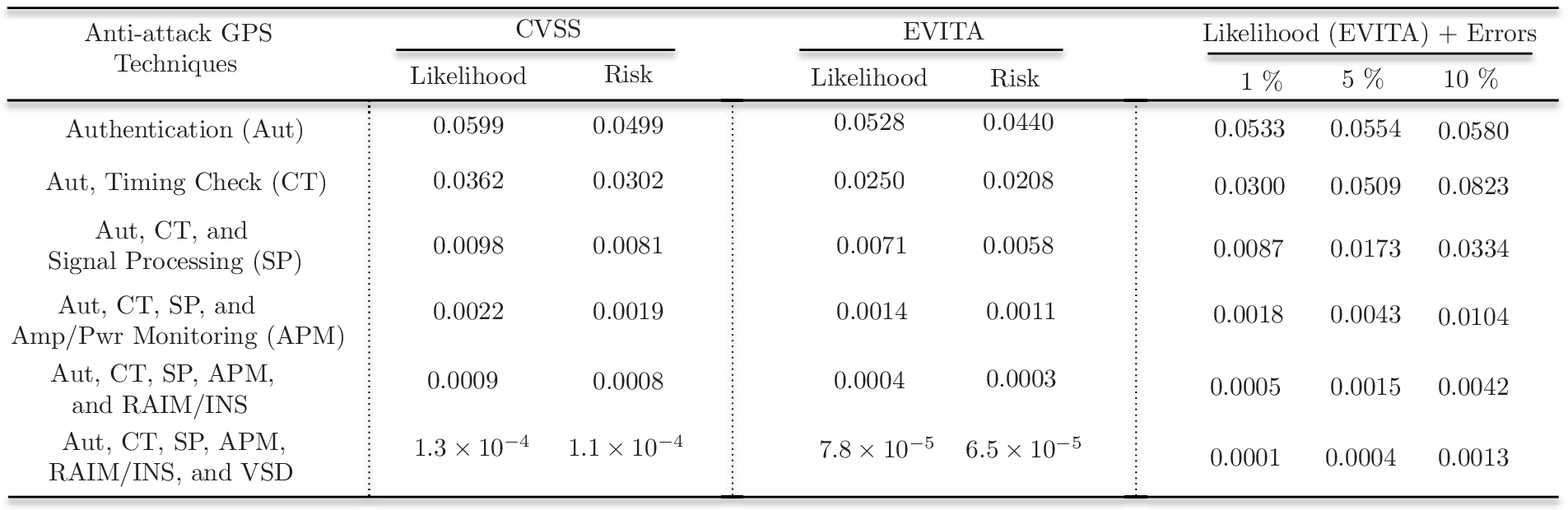}}\label{fuu}
\vspace{-0.07 cm}
\end{table*}

\subsection{Evaluation and Discussion}
In what follows, we evaluate likelihood of threats and risks using equations (\ref{Post}) and (\ref{Risk_std}). To obtain likelihood, we apply Bayesian inference. We initially determine the states of BN model and their roles for detection. It is shown in Fig. \ref{fig:E_c} that there are $16$ nodes, each of which has two states that provide $2^{16}$ possible states. By employing CPTs such as Table \ref{fee}, these states are reduced to $2^6$. Then, we apply equation (\ref{Post}) to obtain the posterior probability of fake GPS signal detection (likelihood) given the incorporated anti-spoofing techniques. Assuming $impact = 0.833$ given by Table \ref{foo}(a) for a GPS component, we can derive the risk defined in  (\ref{Risk_std}). 

Table \ref{fuu} shows resulted beliefs for likelihood and $risk$. Since all the $2^6$ states could not be shown here, a few combinations are selected. It can be seen that the likelihood and the risk of threats are generally decreased by utilizing a higher number of countermeasures. For instance, based on EVITA, the likelihood of attack could be reduced from $5.3 \%$ to less than $0.1 \%$ and $0.01 \%$ by using, respectively, five and six anti-spoofing techniques instead of just one. 
As can be noted, results for CVSS and EVITA are close to each other. This is not surprising, as the prior probabilities of anti-spoofing elements (Table \ref{fii}) are also close. This type of analysis could also help in choosing the number and type of anti-attack techniques to be deployed in the presence of energy, size, and cost limitations. Furthermore, in order to study the resilience of the proposed model, the likelihood of threats is evaluated for different levels of errors. The cause of these errors could vary from noise and inaccurate processing of data to hardware problems in deployed countermeasures. It can be seen that threat likelihood, hence the risk, can be contained to small values, particularly for small errors, when five or more countermeasures are present.


\section{Conclusion}
We have introduced a framework using a Bayesian defense graph to study the cybersecurity of AVs. In particular, we have employed risk assessment models such as EVITA to study the threat likelihood and risk for vulnerable components in AVs in the presence of countermeasures. In a case study, we have applied this framework to infer a belief for the likelihood of threats and risks for GPS signals. Our results confirm that the likelihood of threats can be reduced to $0.01 \%$
depending on what anti-spoofing techniques are employed. Future work will focus on the impact of cooperation between vehicles to improve the security of an AV.

\vspace{+0.4cm}



\bibliographystyle{ieeetran}
\bibliography{biblio1}

\begin{thebibliography}{10}
\providecommand{\url}[1]{#1}
\csname url@samestyle\endcsname
\providecommand{\newblock}{\relax}
\providecommand{\bibinfo}[2]{#2}
\providecommand{\BIBentrySTDinterwordspacing}{\spaceskip=0pt\relax}
\providecommand{\BIBentryALTinterwordstretchfactor}{4}
\providecommand{\BIBentryALTinterwordspacing}{\spaceskip=\fontdimen2\font plus
\BIBentryALTinterwordstretchfactor\fontdimen3\font minus
  \fontdimen4\font\relax}
\providecommand{\BIBforeignlanguage}[2]{{%
\expandafter\ifx\csname l@#1\endcsname\relax
\typeout{** WARNING: IEEEtran.bst: No hyphenation pattern has been}%
\typeout{** loaded for the language `#1'. Using the pattern for}%
\typeout{** the default language instead.}%
\else
\language=\csname l@#1\endcsname
\fi
#2}}
\providecommand{\BIBdecl}{\relax}
\BIBdecl

\bibitem{Petit_Survey}
J.~Petit and S.~E. Shladover, ``Potential cyberattacks on automated vehicles,''
  \emph{IEEE Transactions on Intelligent Transportation Systems}, vol.~16,
  no.~2, pp. 546--556, April 2015.

\bibitem{Survey17}
S.~Parkinson, P.~Ward, K.~Wilson, and J.~Miller, ``Cyber threats facing
  autonomous and connected vehicles: Future challenges,'' \emph{IEEE
  Transactions on Intelligent Transportation Systems}, vol.~18, no.~11, pp.
  2898--2915, Nov 2017.

\bibitem{yan16}
C.~Yan, W.~Xu, and J.~Liu, ``Can you trust autonomous vehicles: Contactless
  attacks against sensors of self-driving vehicle,'' \emph{DEF CON}, vol.~24,
  2016.

\bibitem{FMEA03}
D.~H. Stamatis, \emph{Failure mode and effect analysis: FMEA from theory to
  execution}.\hskip 1em plus 0.5em minus 0.4em\relax ASQ Quality Press, 2003.

\bibitem{petit2015remote}
J.~Petit, B.~Stottelaar, M.~Feiri, and F.~Kargl, ``Remote attacks on automated
  vehicles sensors: Experiments on camera and lidar,'' \emph{Black Hat Europe},
  vol.~11, 2015.

\bibitem{kordy14}
B.~Kordy, L.~Pi{\`e}tre-Cambac{\'e}d{\`e}s, and P.~Schweitzer, ``Dag-based
  attack and defense modeling: Don't miss the forest for the attack trees,''
  \emph{Computer science review}, vol.~13, pp. 1--38, 2014.

\bibitem{roy12}
A.~Roy, D.~S. Kim, and K.~S. Trivedi, ``Attack countermeasure trees (act):
  towards unifying the constructs of attack and defense trees,'' \emph{Security
  and Communication Networks}, vol.~5, no.~8, pp. 929--943, 2012.

\bibitem{kordy10}
B.~Kordy, S.~Mauw, S.~Radomirovi{\'c}, and P.~Schweitzer, ``Foundations of
  attack--defense trees,'' in \emph{International Workshop on Formal Aspects in
  Security and Trust}.\hskip 1em plus 0.5em minus 0.4em\relax Springer, 2010,
  pp. 80--95.

\bibitem{sahara}
G.~Macher, H.~Sporer, R.~Berlach, E.~Armengaud, and C.~Kreiner, ``Sahara: a
  security-aware hazard and risk analysis method,'' in \emph{Design, Automation
  \& Test in Europe Conference \& Exhibition (DATE)}, 2015, pp. 621--624.

\bibitem{islam16}
M.~M. Islam, A.~Lautenbach, C.~Sandberg, and T.~Olovsson, ``A risk assessment
  framework for automotive embedded systems,'' in \emph{Proceedings of the 2nd
  ACM International Workshop on Cyber-Physical System Security}, 2016, pp.
  3--14.

\bibitem{Hu09}
O.~Henniger, L.~Apvrille, A.~Fuchs, Y.~Roudier, A.~Ruddle, and B.~Weyl,
  ``Security requirements for automotive on-board networks,'' in \emph{9th
  International Conference on Intelligent Transport Systems
  Telecommunications,(ITST)}, 2009, pp. 641--646.

\bibitem{ReviewGPS}
A.~Jafarnia-Jahromi, A.~Broumandan, J.~Nielsen, and G.~Lachapelle, ``{GPS}
  vulnerability to spoofing threats and a review of antispoofing techniques,''
  \emph{International Journal of Navigation and Observation}, vol. 2012, no.
  127072, pp. 1--16, July 2012.

\bibitem{EPFL08}
P.~Papadimitratos and A.~Jovanovic, ``{GNSS}-based positioning: Attacks and
  countermeasures,'' in \emph{IEEE Military Communications Conference
  (MILCOM)}, 2008, pp. 1--7.

\bibitem{Mul-Ant15}
J.~Magiera and R.~Katulski, ``Detection and mitigation of {GPS} spoofing based
  on antenna array processing,'' \emph{Journal of Applied Research and
  Technology}, vol.~13, no.~1, pp. 45--57, 2015.

\bibitem{part2}
G.~W. Hein, F.~Kneissi, J.-A. Avila-Rodriguez, and S.~Wallner, ``Authenticating
  {GNSS} proofs against spoofs,'' \emph{Inside GNSS}, pp. 71--78,
  September/October 2007.

\bibitem{VSD08}
T.~E. Humphreys, B.~M. Ledvina, M.~L. Psiaki, B.~W. O'Hanlon, and P.~M.
  Kintner~Jr, ``Assessing the spoofing threat: Development of a portable {GPS}
  civilian spoofer,'' in \emph{Proceedings of the ION GNSS International
  Technical Meeting of the Satellite Division}, vol.~55, 2008, p.~56.

\bibitem{warner03}
J.~S. Warner and R.~G. Johnston, ``{GPS} spoofing countermeasures,''
  \emph{Homeland Security Journal}, vol.~25, no.~2, pp. 19--27, 2003.

\bibitem{psiaki16}
M.~L. Psiaki and T.~E. Humphreys, ``{GNSS} spoofing and detection,''
  \emph{Proceedings of the IEEE}, vol. 104, no.~6, pp. 1258--1270, 2016.

\bibitem{frigault08}
M.~Frigault, L.~Wang, A.~Singhal, and S.~Jajodia, ``Measuring network security
  using dynamic bayesian network,'' in \emph{Proceedings of the 4th ACM
  workshop on Quality of protection}, 2008, pp. 23--30.

\end{thebibliography}


\end{document}